\documentclass[prd,floatfix,preprintnumbers]{revtex4}
\usepackage{amsmath}
\usepackage{graphicx}
\usepackage{dcolumn}
\usepackage{bm}
\usepackage{physics}

\newcommand {\ga} {\ {\raise-.5ex\hbox{$\buildrel>\over\sim$}}\ }
\newcommand {\la} {\ {\raise-.5ex\hbox{$\buildrel<\over\sim$}}\ }

\def\be{\begin{equation}}
\def\ee{\end{equation}}
\def\ba{\begin{eqnarray}}
\def\ea{\end{eqnarray}}

\begin{document}
\title{The Swampland Conjectures and Slow-Roll Thawing Quintessence}
\author{S. David Storm and Robert J. Scherrer}
\affiliation{Department of Physics and Astronomy, Vanderbilt University,
	Nashville, TN  ~~37235}

\begin{abstract}
	
	We examine the Swampland conjectures in the context of generic slow-roll thawing quintessence models.  Defining
		$\lambda \equiv |V^{\prime}(\phi_i)/V(\phi_i)|$ and
	$K \equiv \sqrt{1 - 4V^{\prime \prime}(\phi_i)/3V(\phi_i)}$,
	where $\phi_i$ is the initial value of $\phi$,
	we find regions of parameter space consistent with
	both observational data and with the refined de Sitter conjecture, and we show that all such models satisfy the distance conjecture.  We quantify
	the degree of fine-tuning on $\lambda$ needed to achieve these
	results.
\end{abstract}

\maketitle

Observational evidence \cite{union08,hicken,Amanullah,Union2,Hinshaw,Ade,Betoule}
suggests that approximately
70\% of the energy density in the
universe is in the form of a negative-pressure component called dark energy, with the remaining
30\% in the form of nonrelativistic matter.
The dark energy component can be parametrized in terms of its equation of state parameter, $w$,
defined as the ratio of the dark energy pressure to its density:
\be
\label{w}
w=p/\rho.
\ee
Then a cosmological constant, $\Lambda$, corresponds to the case $\rho = constant$ and $w = -1$ .

While a model with a cosmological constant and cold dark matter ($\Lambda$CDM) is consistent
with current observations,
there are other models of dark energy that have a dynamical equation
of state.
The most widely-investigated are
quintessence models, with a time-dependent scalar field, $\phi$,
having potential $V(\phi)$
\cite{RatraPeebles,Wetterich,Ferreira,CLW,CaldwellDaveSteinhardt,Liddle,SteinhardtWangZlatev}.
(See Ref. \cite{Copeland1} for a review).
In these models, 
the equation of motion for a scalar field is given by
\begin{equation}
\label{phievol}
\ddot{\phi} + 3H\dot{\phi} + \frac{dV}{d{\phi}} = 0,
\end{equation}
where the Hubble parameter H is given by
\begin{equation}
H \equiv \frac{\dot{a}}{a} = \sqrt{\rho_T/3}.
\end{equation}
In this equation, $a$ is the scale factor, $\rho_T$ is the total density, and we take $8 \pi G = \hbar = c = 1$ throughout.
The pressure and density of $\phi$ are given by
\begin{equation}
p_\phi = \frac{\dot{\phi}^2}{2} - V(\phi),
\end{equation}
and 
\begin{equation}
\rho_\phi = \frac{\dot{\phi}^2}{2} + V(\phi). 
\end{equation}

In general, one can simply integrate Eq. (\ref{phievol})
for any given $V(\phi)$, determine the corresponding
$\rho_\phi(a)$, and compare with observational constraints.
More recently, however, a number of authors have also considered the
consistency of quintessence models with a variety
of ``Swampland conjectures."
The Sampland conjectures
arise in the context of attempts to derive a quantum theory of gravity within string theory.
A variety of these conjectures have
been proposed (see, e.g., Refs. \cite{Arkani,OV} for some of the earliest work in this area); we will consider two of them in the context of this paper.

The distance conjecture
constrains the field excursion $\Delta \phi$ to be small when expressed in Planck units, namely

\noindent{\bf Conjecture 1:}
\be
\label{swamp1}
\Delta \phi \la \mathcal{O}(1).
\ee
The distance conjecture is longstanding \cite{OV} and has a great deal
of theoretical support (although see Ref. \cite{Scalisi} for mechanisms to evade it).

More recently, Obied et al. \cite{Obied} proposed  the condition:

\noindent{\bf Conjecture 2.1:}
\be
\label{swamp2.1}
\lambda \equiv |V^\prime/V| \ga \mathcal{O}(1),
\ee
\noindent where $V^\prime$ is the derivative of $V$ with respect to $\phi$.  This constraint, which is based on the difficulty of constructing a de Sitter vacuum in string theory, is called the de Sitter conjecture.

The consistency (or lack thereof) of these conjectures with
quintessence models has been examined in detail
\cite{Brandenberger,limit1,Akrami,Raveri,Garg,Scherrer,Montefalcone,Colgain,Banerjee}.  
(See also Ref. \cite{Cicoli} for multi-field quintessence, and Refs. \cite{Brahma1,Trodden,Brahma2} for scalar
fields beyond quintessence). In general,
it is extremely difficult to reconcile the de Sitter conjecture
with standard quintessence models (and it is inconsistent
with standard $\Lambda$CDM).
The problem arises because observations favor $w$ near $-1$ at moderate redshifts, but $w$ near $-1$ generally translates
into values of $\lambda$ less than 1.  (See Ref. \cite{Geng} for
a mechanism to evade this issue in the context of inflation).

Subsequently, an alternative de Sitter conjecture, called the refined de Sitter conjecture, was proposed \cite{Garg,OPSV}.  This condition is
that the scalar field satisfies {\it either}
Eq. (\ref{swamp2.1}) or the following:

\noindent{\bf Conjecture 2.2:}
\be
\label{swamp2.2}
c^2 \equiv - V^{\prime\prime}/V \ga  \mathcal{O}(1).
\ee
Thus, Eq. (\ref{swamp2.2}) requires the potential $V(\phi)$ to be concave, with a lower bound
on the curvature.

Eq. (\ref{swamp2.2}) leads to a natural quintessence model, namely, one in which the scalar field begins
initally near a maximum in the potential and then rolls
downhill.
Such ``hilltop" quintessence models have been considered in some detail.  In the terminology of Ref. \cite{CL},
these are ``thawing" models, in which the field begins initially with $\dot \phi \approx 0$ at some initial value $\phi = \phi_i$.
Initially, $w \approx -1$, but
the field then evolves with $\dot \phi$ increasing with time up to the present, so that $w$ increases
with time.  Note that in these models, any initial nonzero
value of $\dot \phi$ is damped by Hubble friction (the
second term in equation \ref{phievol}) at early times, so the assumption
that $\dot \phi_i \approx 0$ is reasonable.

While quintessence generically produces a time-varying value for
$w$, 
a successful model must closely mimic $\Lambda$CDM in order to be consistent
with current observations. Hence, a viable model should yield a present-day value
of $w$ close to $-1$.
This fact has been exploited in a number of papers that explored the evolution of a scalar field
subject to the constraint that $w$ must be close to $-1$
\cite{ds1,Chiba,ds2}.  By imposing this constraint, one can reduce an
infinite number of models to a finite set of behaviors for $w(a)$ characterized entirely by the values of
$V^\prime/V$ and $V^{\prime \prime}/V$ at the initial value
of the scalar field, $\phi_i$. Here we will use this methodology 
to determine the parameter space consistent with current observations and allowed by the distance conjecture and the refined de Sitter conjecture.

Constraints on quintessence models from the refined de Sitter conjecture have been examined previously by Agrawal and Obied \cite{AO} and by Raveri et al. \cite{Raveri}.  The former examined a potential consisting
of a constant plus a negative quadratic, while the latter
used a potential of the form 
\be
\label{Raveripotential}
V(\phi) = B \cos(c \phi).
\ee
Our own investigation most closely resembles that in
Ref. \cite{Raveri}, so we will place our own results
in the context of that paper.  Raveri et al. allowed for a variation in $B$, $c$, and $\phi_i$ (the initial value of $\phi$), and their results indicate that $c>1$ is already disfavored by current observational constraints.  Of course, it is obvious (and noted in Ref. \cite{Raveri}) that arbitrarily large values of $c$ are possible if $\phi_i$ is sufficiently small, but such conditions require a high degree of fine-tuning and can be destabilized by quantum fluctuations.

Our approach differs somewhat from that in Ref. \cite{Raveri}.  We use the formalism of Refs. \cite{ds1,Chiba,ds2} to examine arbitrary thawing models with $w$ near $-1$.  While it is possible to map the parameter space of Ref. \cite{Raveri} onto this formalism, our approach corresponds to a different set of priors on these parameters, which will result in a different set of constraints.  Further, our approach leads naturally to a useful relation between $\Delta \phi$ in the distance conjecture and $V^{\prime \prime}/V$ in the refined de Sitter conjecture.  Finally, we explore the exact degree of fine-tuning needed in models that satisfy the refined de Sitter conjecture for generic thawing models. 

Consider a thawing scalar field with a potential $V(\phi)$
characterized by $\lambda \equiv |V^\prime/V| < 1$.  Following Refs. \cite{ds1,Chiba,ds2}, we note that such a
potential will lead naturally to $1 + w \ll 1$ at present,
and the evolution of $\phi$ and $w$ will follow
a well-defined, limited set of trajectories.  Note that
our assumption violates Eq. (\ref{swamp2.1}), so that for
the models considered here, the refined Swampland conjecture
will require Eq. (\ref{swamp2.2}) to be satisfied.
The evolution of $\phi$ as a function of $t$ in this case is \cite{Chiba}
\begin{equation}
\label{phi(t)}
\phi(t) = \phi_i + \frac{V'(\phi_i)}{V''(\phi_i)}\Biggr( \frac{\sinh(kt)}{k t_\Lambda \sinh({t}/{t_\Lambda})} -1 \Biggr),
\end{equation}
where $\phi_i$ is the initial value of $\phi$, $t$ is the time, $t_\Lambda$ is the value of $t$ given by
\begin{equation}
t_\Lambda = \frac{2}{\sqrt{3V(\phi_i)}},
\end{equation}
and the constant $k$ characterizes the curvature of the potential:
\begin{equation}
k\equiv\sqrt{\frac{3V(\phi_i)}{4} -  V''(\phi_i)},
\end{equation}

We can use Eq. (\ref{phi(t)}) to find the field
excursion $\Delta \phi$ between $t = 0$ and $t = t_0$
(the present time).  Taking the background expansion
to be approximately $\Lambda$CDM, we have
\begin{equation}
\label{t0}
t_0 = t_\Lambda \tanh[-1](\sqrt{\Omega_{\phi 0}}).
\end{equation}
and following Refs. \cite{ds1,Chiba,ds2}, we define
\begin{equation}\label{eq:K}
K \equiv kt_\Lambda = \sqrt{1- \frac{4V''(\phi_i)}{3V(\phi_i)}}.
\end{equation}
Then Eq. (\ref{phi(t)}) allows us to express the field
excursion as
	\begin{equation}\label{eq:main1}
\mid\Delta\phi\mid = \abs{\frac{4\lambda}{3(1-K^2)}\left( \frac{(1-\Omega_{\phi 0})^{\frac{1}{2}(1-K)} \Bigr[(1+\sqrt{\Omega_{\phi 0}})^K - (1-\sqrt{\Omega_{\phi 0}})^K \Bigr]}{2K\sqrt{\Omega_{\phi 0}}} - 1 \right)}.
\end{equation}
Note that in the limit where $K \rightarrow 1$ ($V^{\prime \prime} \rightarrow 0$), we regain the expression in
Ref. \cite{Raveri}
for $|\Delta \phi|$ when $V^\prime(\phi) =$ const.
In Fig. 1, we show the relationship between
$|\Delta \phi|$, $\lambda$, and $K$ given by Eq. (\ref{eq:main1}). (We take $\Omega_{\phi 0} = 0.69$ throughout).
\begin{figure}[ht]
\includegraphics[width=\linewidth]{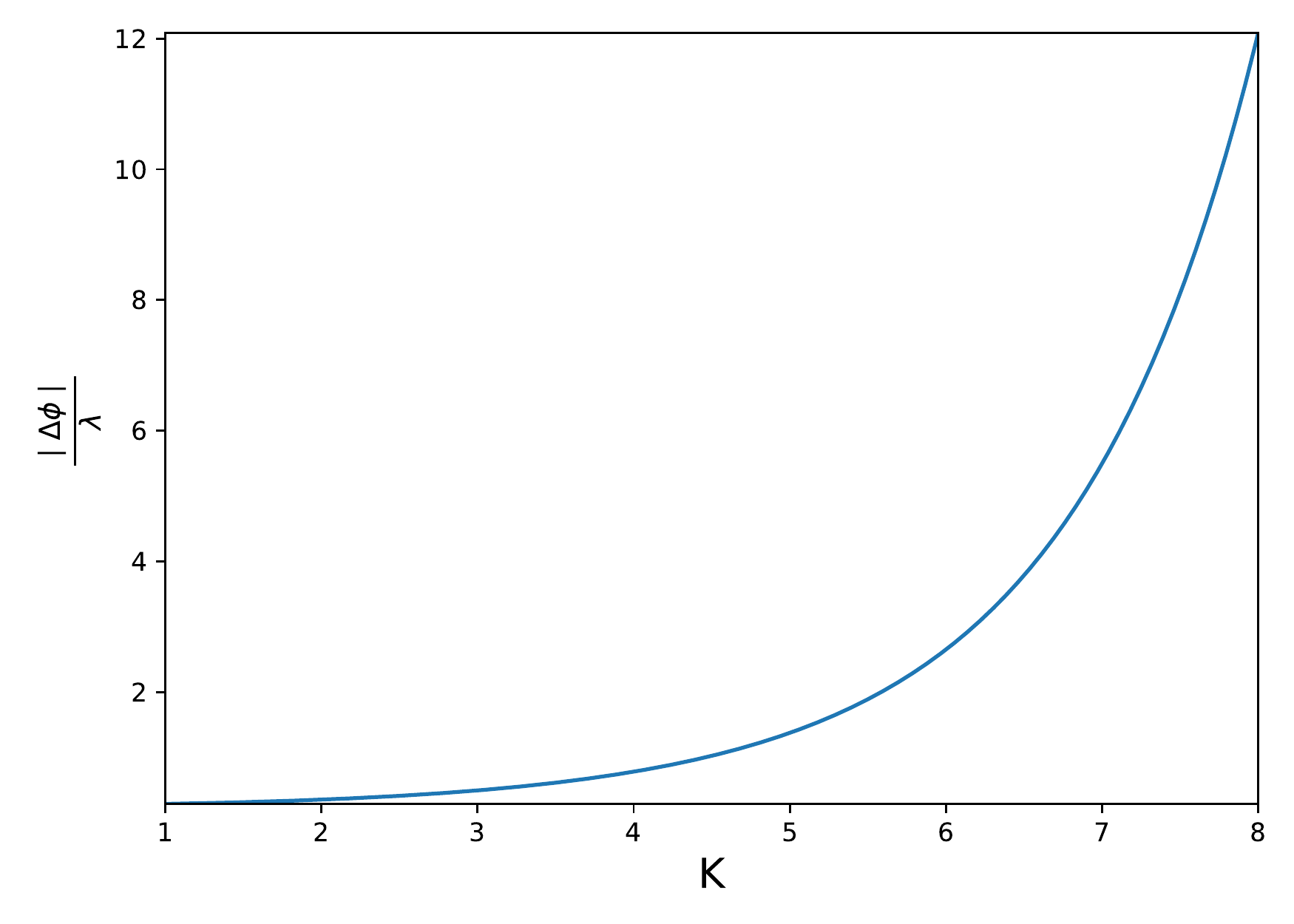}
\caption{For slow-roll thawing quintessence, the field excursion $|\Delta \phi|$ divided by
	$\lambda \equiv |V^{\prime}(\phi_i)/V(\phi_i)|$ as a function of
	$K \equiv \sqrt{1 - 4V^{\prime \prime}(\phi_i)/3V(\phi_i)}$, where $\phi_i$ is the initial value of $\phi$.}
\label{fig:main1}
\end{figure}

Now we will solve for $w_0$, the present-day value of $w$, as a function of $\lambda$ and $K$.  From
Ref. \cite{Chiba}, we have
\begin{equation} \label{eq:chiba26}
1+w(t) = \frac{3}{4}\Biggr( \frac{V'(\phi_i)}{kt_\Lambda V''(\phi_i)} \Biggr)^2 \Biggr(\frac{kt_\Lambda \cosh(kt)\sinh({t}/{t_\Lambda}) - \sinh(kt)\cosh({t}/{t_\Lambda})}{\sinh[2]({t}/{t_\Lambda})} \Biggr)^2
\end{equation}
Taking $t$ equal to the present-day time from Eq. (\ref{t0}),
we obtain
\begin{equation} \label{eq:main2}
1+w_0 = \frac{\lambda^2}{3K^2(1-K^2)^2}\frac{(1-\Omega_{\phi 0})^{1-K}}{\Omega_{\phi 0}}\Biggr[ (K-\Omega_{\phi 0}^{-\frac{1}{2}})(1+\sqrt{\Omega_{\phi 0}})^K + (K+\Omega_{\phi 0}^{-\frac{1}{2}})(1-\sqrt{\Omega_{\phi 0}})^K \Biggr]^2 .
\end{equation}
This result is illustrated in Fig. 2.
\begin{figure}[ht]
	\includegraphics[width=\linewidth]{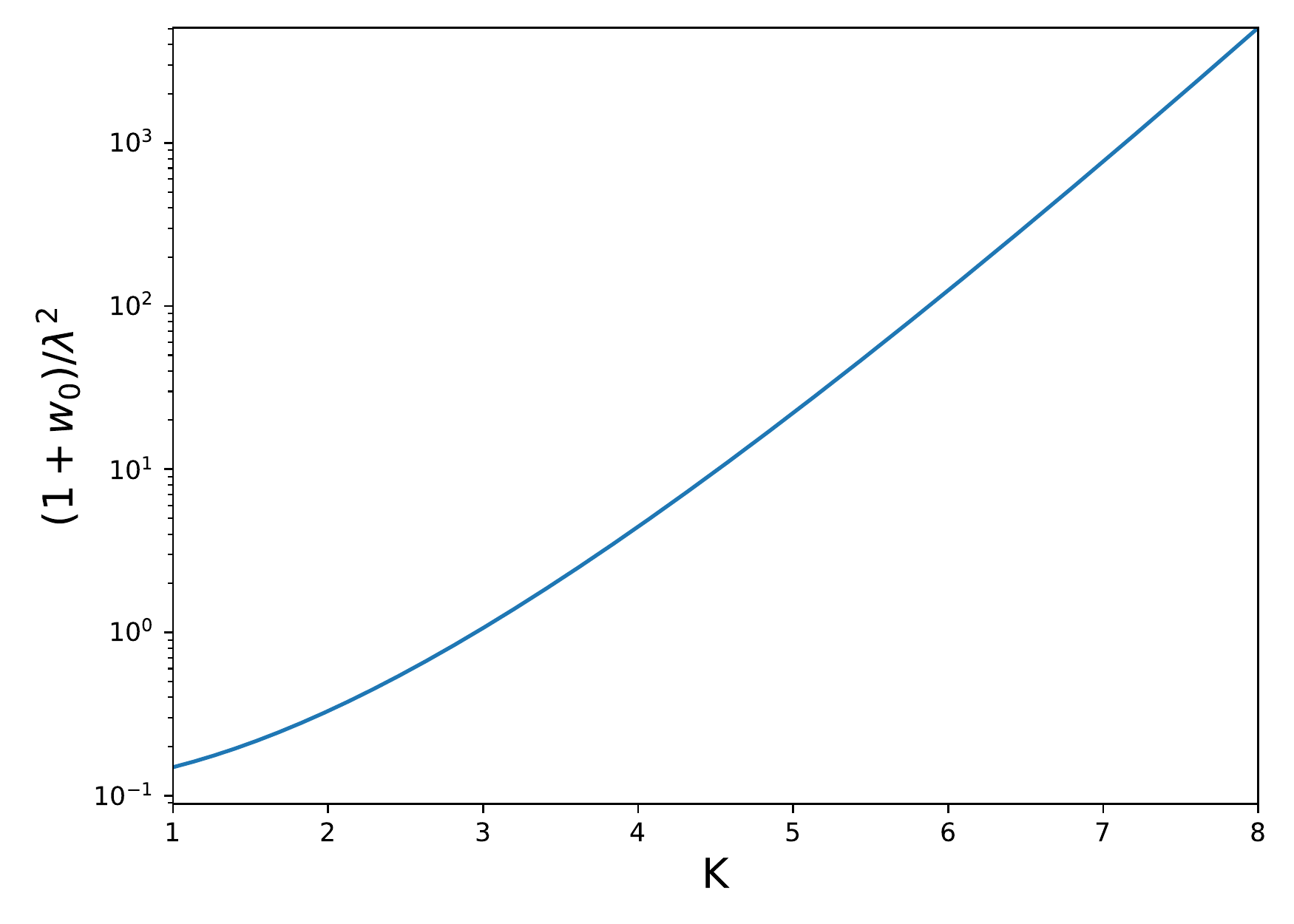}
	\caption{For slow-roll thawing quintessence, the
		present day value of the equation of state parameter,
		$1+w_0$, divided by
		$\lambda \equiv |V^{\prime}(\phi_i)/V(\phi_i)|$ as a function of
		$K \equiv \sqrt{1 - 4V^{\prime \prime}(\phi_i)/3V(\phi_i)}$, where $\phi_i$ is the initial value of $\phi$.}
	\label{fig:main2}
\end{figure}
In the limit where $K \rightarrow 1$ ($V^{\prime \prime} \rightarrow 0$), we obtain the corresponding expression
for $(1+w_0)/\lambda^2$ for a linear potential \cite{ScherrerSen}.

\begin{figure}[ht]
	\includegraphics[width=\linewidth]{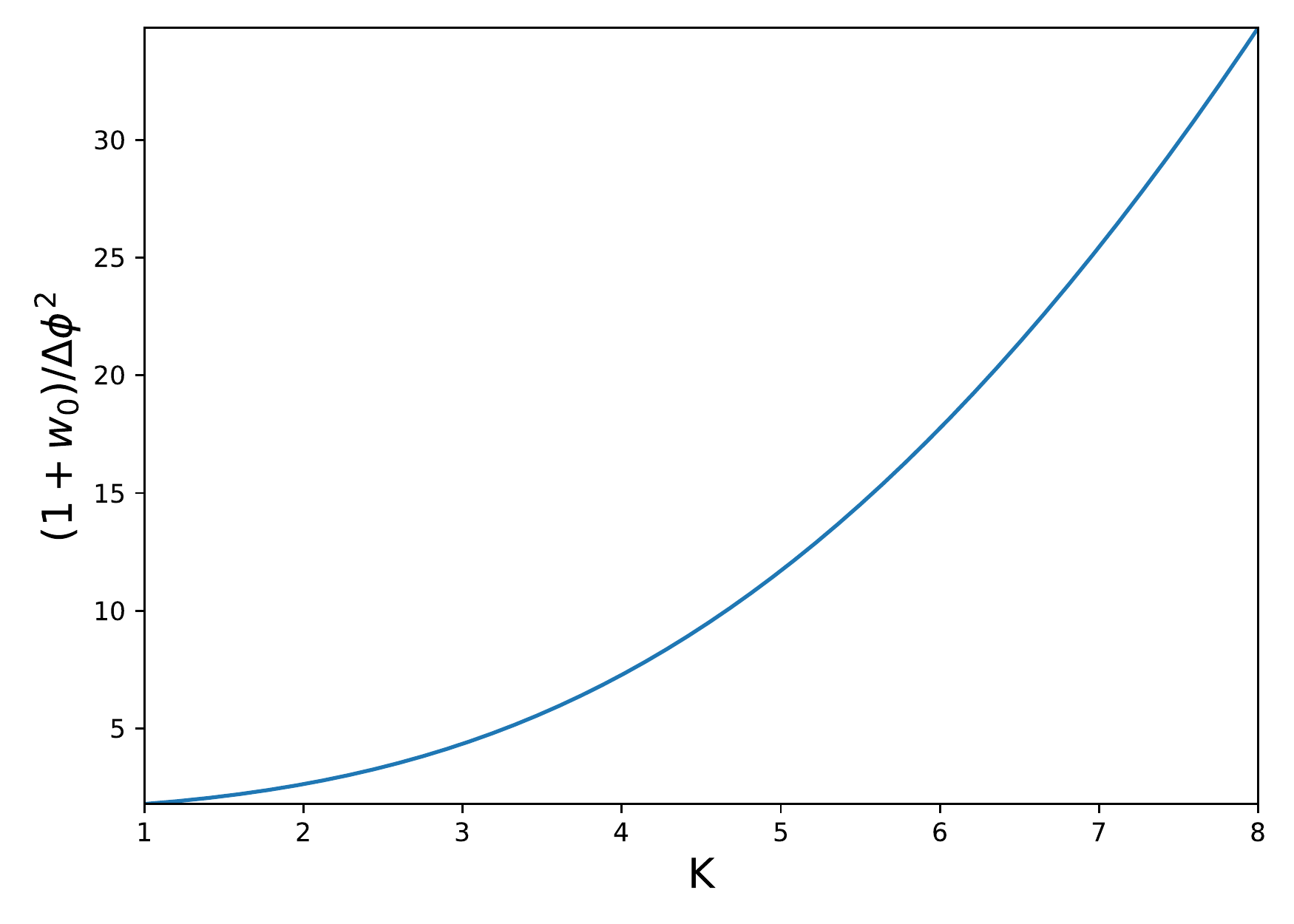}
	\caption{For slow-roll thawing quintessence, the
		present day value of the equation of state parameter,
		$1+w_0$, divided by
		the square of the field excursion, $\Delta \phi$, as a function of
		$K \equiv \sqrt{1 - 4V^{\prime \prime}(\phi_i)/3V(\phi_i)}$, where $\phi_i$ is the initial value of $\phi$.}
	\label{fig:main3}
\end{figure}

Finally, we can combine Eqs. (\ref{eq:main1}) and (\ref{eq:main2}) to eliminate $\lambda$, giving
\begin{equation} \label{eq:main3}
	1+w_0 = \frac{3}{4}\Delta\phi^2 \Biggr[ \frac{(K-\Omega_{\phi 0}^{-\frac{1}{2}})(1+\sqrt{\Omega_{\phi 0}})^K  + (K+\Omega_{\phi 0}^{-\frac{1}{2}})(1-\sqrt{\Omega_{\phi 0}})^K}{(1+\sqrt{\Omega_{\phi 0}})^K - (1-\sqrt{\Omega_{\phi 0}})^K - 2K\sqrt{\frac{\Omega_{\phi 0}}{1-\Omega_{\phi 0}}}(1-\Omega_{\phi 0})^{\frac{K}{2}}} \Biggr]^2 .
	\end{equation}
This result is illustrated in Fig. (\ref{fig:main3}) .

Now consider the constraints that the refined Swampland conjecture (in the form of Eq. \ref{swamp2.2}) places on these slow-roll
thawing models when combined with observational constraints.  Observational limits on slow-roll thawing
models have been considered by Chiba, et. al. \cite{CDT}, Smer-Berreto and
Liddle \cite{SL}, and Durrive, et al. \cite{durrive2018}.
Of these papers, the treatment in Refs. \cite{CDT} and \cite{durrive2018} is most similar to the discussion here.
Both of these papers use the methodology of Refs. \cite{ds1,Chiba,ds2}, along with observational data,
to provide constraints in the $K, w_0$ plane.
Ref. \cite{SL} takes a somewhat different approach, deriving
observational constraints on the Pseudo-Nambu-Goldstone Boson (PNGB) potential
\be
\label{PNGB}
V(\phi) = M^4 \left[1 + \cos(\phi/f)\right],
\ee
and then converting those constraints into limits on $K$ and $w_0$.  Here we will use the limits derived in Ref.
\cite{durrive2018}, but we will return to the approach in
Ref. \cite{SL} at the end of the paper.

Ref. \cite{durrive2018} derives limits 
on the slow-roll thawing models
using the Planck 2015 release \cite{Planck}, type Ia supernova observations from SDSS-II and SNLS \cite{Betoule},
and baryon acoustic oscillation (BAO) measurements \cite{BAO1,BAO2,BAO3}.   The resulting $2\sigma$ confidence limits
in the plane defined  by $w_0$ and $K$ can be mapped directly onto the models examined here.
Using Eq.
(\ref{eq:main3}), we can transform these limits
into a corresponding region in the plane defined
by $|\Delta \phi|$ and $K$; this region is presented in
Fig. 4.  (Note that we fix $\Omega_{\phi 0}= 0.69$, while
Durrive et al. marginalize over this quantity.  However,
their corresponding distribution for $\Omega_{\phi 0}$ (marginalized over $w_0$) is strongly peaked
near our value, and Eqs. (\ref{eq:main1}),
(\ref{eq:main2}), and (\ref{eq:main3}) are slowly-varying functions of $\Omega_{\phi0}$, so our results are
relatively insensitive to the assumed value
for $\Omega_{\phi 0}$).
\begin{figure}[ht]
	\includegraphics[width=\linewidth]{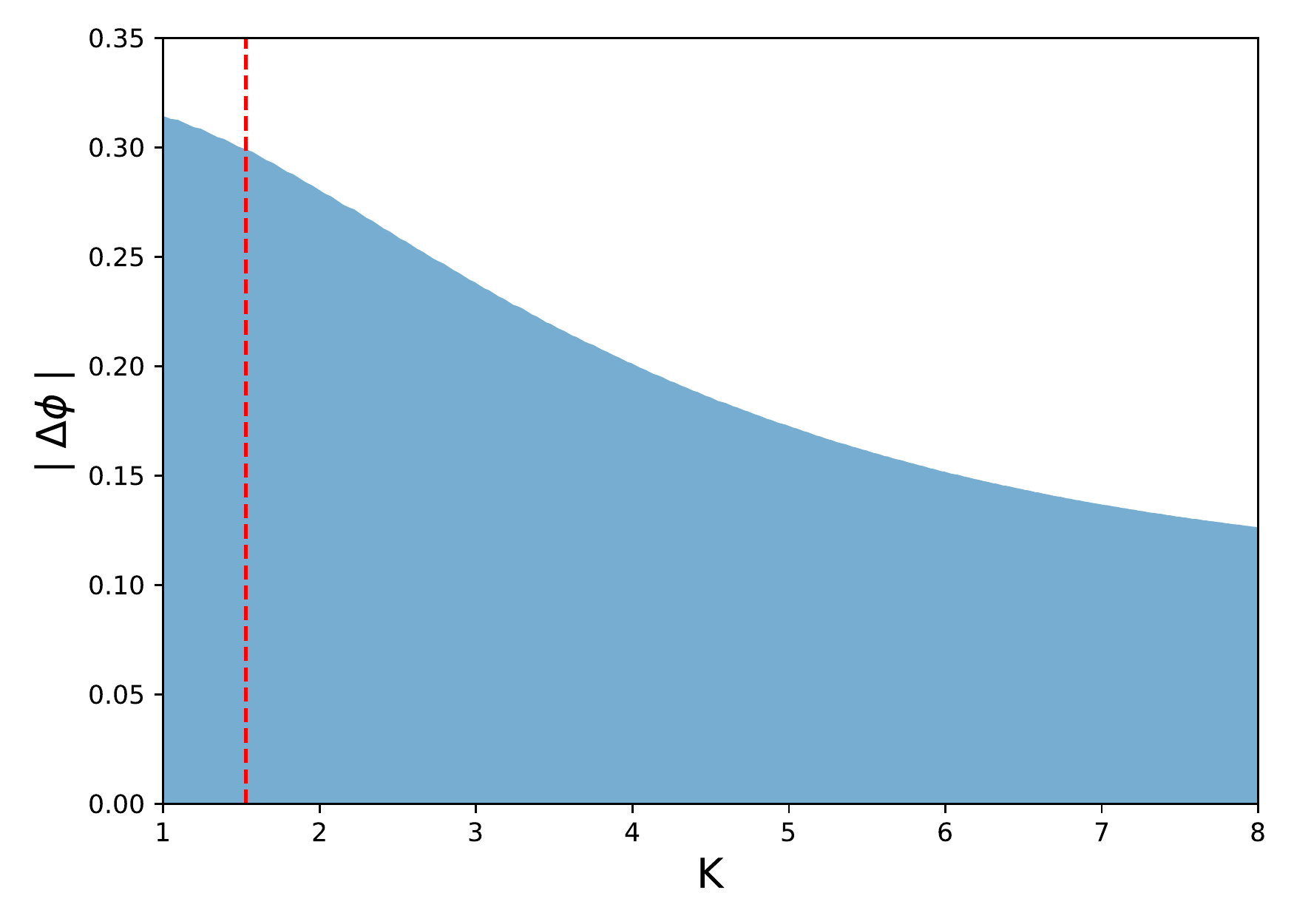}
	\caption{Observational constraints on
		slow-roll thawing quintessence from
		Ref. \cite{durrive2018} mapped onto the
		$|\Delta \phi|$, $K$ plane using the results
		of the present paper, where	$\Delta \phi$ is
		the distance traversed by the field up to the present, and
		$K \equiv \sqrt{1 - 4V^{\prime \prime}(\phi_i)/3V(\phi_i)}$, with $\phi_i$ the initial value of $\phi$.  The area outside of the shaded blue region
		corresponds to the $2\sigma$ excluded region in
		Ref. \cite{durrive2018}. The region to the right of
		the vertical red line satisfies the refined de Sitter
		conjecture with $c>1$. }
	\label{fig:main5}
\end{figure}
If we use the limit $-V^{\prime \prime}/V > 1$ for the
refined de Sitter conjecture, then the region
allowed by this conjecture
lies to the right of the vertical red dashed line:  namely,
$K > \sqrt{7/3} = 1.53$.
From Fig. 4, it is clear that for the entire region satisfying both the observational constraints and the refined de Sitter conjecture, we have
$|\Delta \phi| < 0.3$.  Thus, for slow-roll thawing quintessence models consistent with current observations, the distance conjecture becomes redundant; it is
satisfied whenever the refined de Sitter conjecture is
satisfied, while the converse is not true.

In the same way we can use Eq.
(\ref{eq:main2}) to transform the observational limits
from Ref. \cite{durrive2018}
into a corresponding region in the plane defined
by $\lambda$ and $K$; this region is presented in
Fig. 5. 
\begin{figure}[ht]
	\includegraphics[width=\linewidth]{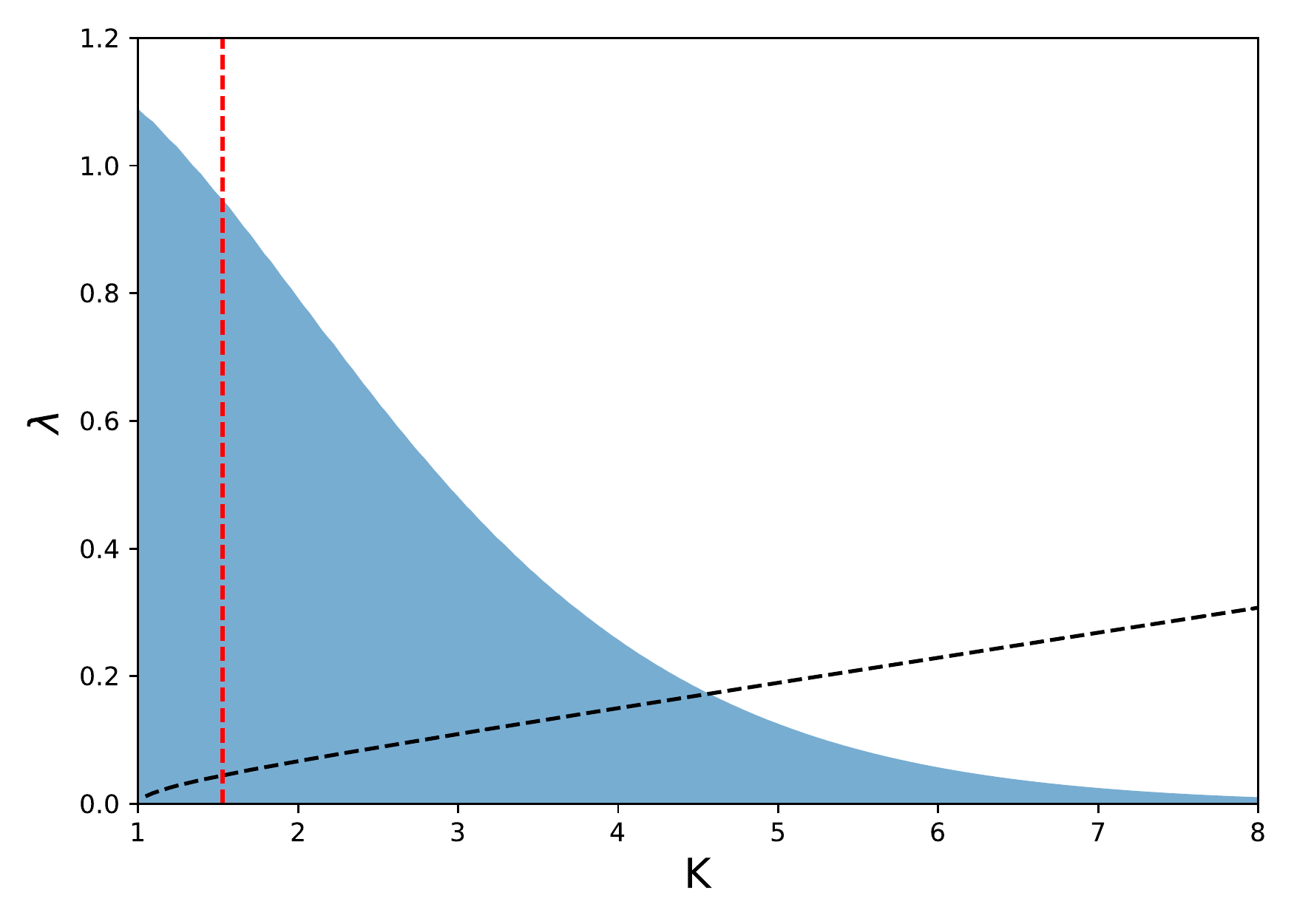}
	\caption{Observational constraints on
		slow-roll thawing quintessence from
		Ref. \cite{durrive2018} mapped onto the
		$\lambda$, $K$ plane using the results
	of the present paper, where	
	$\lambda \equiv |V^{\prime}(\phi_i)/V(\phi_i)|$ and
	$K \equiv \sqrt{1 - 4V^{\prime \prime}(\phi_i)/3V(\phi_i)}$, with $\phi_i$ the initial value of $\phi$.  The area outside of the shaded blue region
	corresponds to the $2\sigma$ excluded region in
	Ref. \cite{durrive2018}. The region to the right of
	the vertical red line satisfies the refined de Sitter
	conjecture with $-V^{\prime \prime}/V>1$.  The region below the black
	curve was excluded from consideration in Ref. \cite{Raveri} (see text).}
	\label{fig:main4}
\end{figure}
As expected, one can find initial conditions consistent
with observations for arbitrarily large values of $K$; however, as the curvature
becomes larger, the value of $\lambda$ must become increasingly finely tuned to smaller values.  This
corresponds to placing the field initially closer
to a local maximum in the potential.  Obviously, these finely-tuned initial conditions are less plausible, but the question of what actually defines fine-tuning is, to some extent, arbitrary.  However, we can take advantage of the form of the allowed region to make some plausible statements about fine tuning.  For the region
\be
\label{allowedK}
\sqrt{7/3} < K < 5,
\ee the corresponding value of $\lambda$ is $\sim 0.1 - 1$, which would certainly not
be considered fine tuned.  On the other hand, for $K > 6$,
we have $\lambda < 0.05$, with $\lambda$ approaching zero very rapidly as $K$ increases beyond this value.  Hence, a plausible region consistent with observations, allowed by the refined de Sitter conjecture, and not particularly fine tuned is defined by
Eq. (\ref{allowedK}).

This conclusion differs somewhat from that in Raveri et al. \cite{Raveri}, who found a strong tension between the refined de Sitter conjecture and potentials of the form given by
Eq. (\ref{Raveripotential}).  However, these two results are not
inconsistent.  First note that Ref. \cite{Raveri} used the cutoff $c \phi_i < 0.0447$, which eliminates the smallest values of $\lambda$.  This cutoff corresponds to the region below the black curve in Fig. 5.  We have discounted some of this region already as corresponding to fine-tuned values of $\lambda$, but this does not account for our large allowed region with $K < 5$.  This region
is downweighted in Ref. \cite{Raveri} because these authors
use a flat prior on the quantity $B/\rho_{\phi 0}$, which disfavors larger values of $c$ (corresponding to larger $K$
in our discussion.) A similar effect can be seen in comparing
the observational constraints (without Swampland considerations) in Refs. \cite{CDT} and \cite{durrive2018} to those in Ref.
\cite{SL}.  Both \cite{CDT} and \cite{durrive2018} find
broad observationally-allowed regions in the $1+w_0$, $K$ plane.
Ref. \cite{SL}, on the other hand, adopts an approach similar to
that of Raveri et al., beginning with a PNGB potential
(Eq. \ref{PNGB}) and deriving a corresponding excluded region
for $K$ and $w_0$.  The prior on $K$ in this case is highly
non-uniform, resulting in a much smaller allowed region.
Indeed, if we were to use the observational constraints of Ref. \cite{SL} and
repeat our analysis here, our conclusions would be much closer
to those of Ref. \cite{Raveri}.  Neither approach is intrinsically more correct; the difference simply illustrates that the constraints
on quintessence models are very sensitive
to the assumed priors.

The results presented here by no means exhaust the various Swampland conjectures that have been proposed.  For example, Bedroya and Vafa \cite{BV}
have recently proposed the ``trans-Planckian censorship conjecture," with the corresponding cosmological consequences examined in Ref. \cite{HBBR}.
However, the formalism we have utilized here for slow-roll thawing quintessence is particularly well suited to examining the refined de Sitter conjecture in the form of Eq. (\ref{swamp2.2}), as this formalism naturally accounts for the behavior of scalar fields
near the maximum of potentials with negative curvature.

\section*{Acknowledgments}

R.J.S. was supported in part by the Department of Energy (DE-SC0019207).  We thank Andrew Liddle for helpful discussions.

\end{document}